%% file: main.tex
\documentclass[amsmath,amssymb,aps,prl,superscriptaddress,twocolumn,floatfix]{revtex4-1}

\usepackage{graphicx}
\usepackage{hyperref}
\usepackage{textcomp} 
\usepackage[english]{babel}
\usepackage{siunitx}
\usepackage{nicefrac}
\usepackage{xcolor}
\usepackage{microtype}

\DeclareSIUnit\gauss{G}
\usepackage{blindtext}
\hypersetup{colorlinks=true, linkcolor=blue,citecolor=blue,urlcolor=blue}

\newcommand{\myref}[2][]{\hyperref[#2]{Fig.~\ref*{#2}#1}}
\newcommand{\Myref}[2][]{\hyperref[#2]{Figure~\ref*{#2}#1}}
\newcommand{\Mytabref}[2][]{\hyperref[#2]{Table~\ref*{#2}#1}}
\newcommand{\mytabref}[2][]{\hyperref[#2]{Tab.~\ref*{#2}#1}}

\begin{document}

\preprint{ErDyMOT}

\author{P. Ilzh\"ofer}
\affiliation{%
 Institut f\"{u}r Experimentalphysik und Zentrum f\"{u}r Quantenoptik,\\ Universit\"{a}t Innsbruck, Technikerstrasse 25, 6020 Innsbruck, Austria
}%
\affiliation{%
 Institut f\"{u}r Quantenoptik und Quanteninformation, \"Osterreichische Akademie der Wissenschaften, 6020 Innsbruck, Austria
}%
\author{G.~Durastante}
\affiliation{%
 Institut f\"{u}r Experimentalphysik und Zentrum f\"{u}r Quantenoptik,\\ Universit\"{a}t Innsbruck, Technikerstrasse 25, 6020 Innsbruck, Austria
}%
\affiliation{%
 Institut f\"{u}r Quantenoptik und Quanteninformation, \"Osterreichische Akademie der Wissenschaften, 6020 Innsbruck, Austria
}%
\author{A.~Patscheider}
\affiliation{%
 Institut f\"{u}r Experimentalphysik und Zentrum f\"{u}r Quantenoptik,\\ Universit\"{a}t Innsbruck, Technikerstrasse 25, 6020 Innsbruck, Austria
}%
\affiliation{%
 Institut f\"{u}r Quantenoptik und Quanteninformation, \"Osterreichische Akademie der Wissenschaften, 6020 Innsbruck, Austria
}%
\author{A.~Trautmann}
\affiliation{%
 Institut f\"{u}r Quantenoptik und Quanteninformation, \"Osterreichische Akademie der Wissenschaften, 6020 Innsbruck, Austria
}%
\author{M.~J.~Mark}
\affiliation{%
 Institut f\"{u}r Experimentalphysik und Zentrum f\"{u}r Quantenoptik,\\ Universit\"{a}t Innsbruck, Technikerstrasse 25, 6020 Innsbruck, Austria
}%
\affiliation{%
 Institut f\"{u}r Quantenoptik und Quanteninformation, \"Osterreichische Akademie der Wissenschaften, 6020 Innsbruck, Austria
}%

\author{F. Ferlaino}
\affiliation{%
 Institut f\"{u}r Experimentalphysik und Zentrum f\"{u}r Quantenoptik,\\ Universit\"{a}t Innsbruck, Technikerstrasse 25, 6020 Innsbruck, Austria
}%
\affiliation{%
 Institut f\"{u}r Quantenoptik und Quanteninformation, \"Osterreichische Akademie der Wissenschaften, 6020 Innsbruck, Austria
}%
\title{
A two-species five-beam magneto-optical trap for highly magnetic Er and Dy atoms
}
\date{November 2017}

\begin{abstract}
We report on the first realization of a two-species magneto-optical trap (MOT) for erbium and dysprosium. The MOT operates on an intercombination line for the respective species. Owing to the narrow-line character of such a cooling transition and the action of gravity, we demonstrate a novel trap geometry employing only five beams in orthogonal configuration. We observe that the mixture is cooled and trapped very efficiently, with up to \num{5e8} Er atoms and \num{e9} Dy atoms at temperatures of about \SI{10}{\micro K}. Our results offer an ideal starting condition for the creation of a dipolar quantum mixture of highly magnetic atoms.
\end{abstract}

\maketitle

Within the very active research field of ultracold quantum gases, heteronuclear mixtures of different atomic species offer unique possibilities to study a broad range of quantum phenomena. In the past 15 years, various atomic species have been combined to produce quantum degenerate mixtures. Each of such quantum mixtures has its own characteristic traits. Among the widely used alkali mixtures, e.\,g.~\cite{Hadzibabic2002,Roati2002}, the mass imbalance and the selective tuning of the intra- and interspecies interaction have allowed to investigate fascinating phenomena, such as heteronuclear Efimov states \cite{Barontini2009,Pires2014, Tung2014, Wacker2016}, polaron and impurity physics in both bosonic and fermionic backgrounds \cite{Schirotzek2009, Koschorreck2012, Kohstall2012, Rentrop2016, Jorgensen2016}, and heteronuclear molecules with large electric dipole moments \cite{Ni2008, Takekoshi2014, Molony2014}. 

The latter development is mainly driven by the interest in studying phenomena arising from the long-range and anisotropic dipole-dipole interactions among the molecules \cite{Pupillo2008}. As an alternative approach, magnetic atoms have proven to be a robust system to study few- and many-body dipolar physics. The strength of magnetic atoms for the study of dipolar physics has been first shown using Bose-Einstein condensates of chromium atoms \cite{Griesmaier2005, Beaufils2008}. More recently, both erbium (Er) and dysprosium (Dy), among the most magnetic and isotope-rich atomic species, have been individually brought to quantum degeneracy \cite{Lu2011,Lu2012, Aikawa2012,Aikawa2014a}. Using these species, remarkable many-body dipolar phenomena have been observed, including the observation of deformed Fermi surfaces \cite{Aikawa2014}, quantum droplets \cite{Kadau2016, Schmitt2016, Chomaz2016}, roton excitations \cite{Chomaz2017}, and the recent study of thermalization in many-body dipolar gases \cite{Yijun2017}.

Adding the flexibility of mixtures to the richness of magnetic atoms, we here report on the first combination of the two highly magnetic atomic species Er and Dy in a single experimental apparatus. The Er-Dy system extends the collection of available quantum mixtures by an unexplored case, as the interplay between the interspecies contact and dipolar interactions, and the dipolar imbalance among the two species provides a new dimension in the parameter space of accessible quantum phenomena. This impacts, e.\,g.,~the miscibility properties of the mixture \cite{Kumar2017}. Despite imbalanced dipolar mixture systems have not yet been considered in theory, they are good candidates to observe, e.\,g.,~long-range dominated polarons, dipolar pairing, and the anisotropic BEC-BCS crossover with deformed Fermi surfaces.

While single-species magneto-optical traps (MOTs) of Er \cite{Berglund2007,Frisch2012,Ulitzsch2017} and Dy \cite{Lu2010,Maier2014,Dreon2017} as well as other lanthanoid atoms \cite{Kuwamoto1999,Sukachev2010,Miao2014} have already been attained, we simultaneously cool and trap Er and Dy in a two-species MOT operating on intercombination lines. We observe a remarkably robust operation of the dual MOT with atom numbers similar or even surpassing the typical ones recorded in the single-species Er or Dy experiments. Moreover, we demonstrate magneto-optical trapping using a unique beam configuration, allowing us to efficiently operate the MOT using only five beams (5B) in an orthogonal \emph{open-top} configuration, see \myref[ (b)]{fig:levelsandsetup}. The working principle of our orthogonal 5B MOT relies on the combined effect of the narrow-line cooling and gravity \cite{Katori1999,Kuwamoto1999,Frisch2012,Maier2014,Ulitzsch2017,Dreon2017}, and contrasts the classical six beam (6B) approach.

\begin{figure*}[ht]
\flushleft
    \includegraphics[trim = 0 90 50 70,clip]{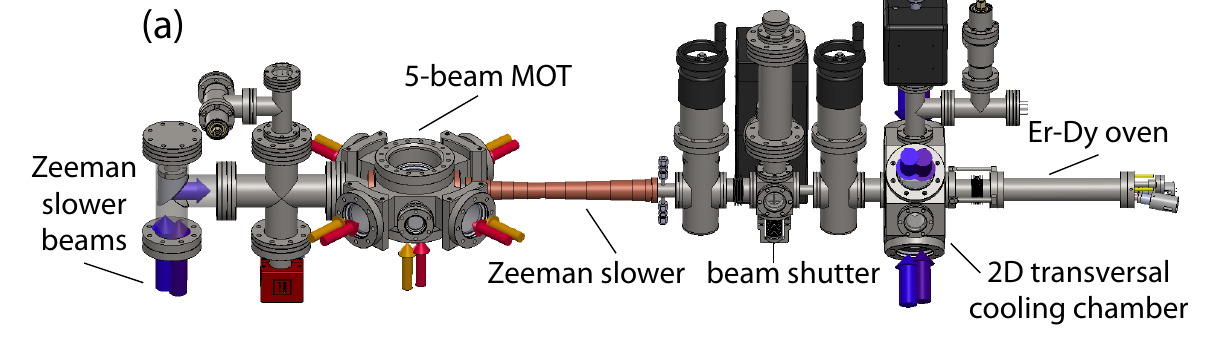}
    \hfill
    \parbox{4cm}{\vspace*{-2.95cm}\hspace*{-1.3cm}
        \includegraphics{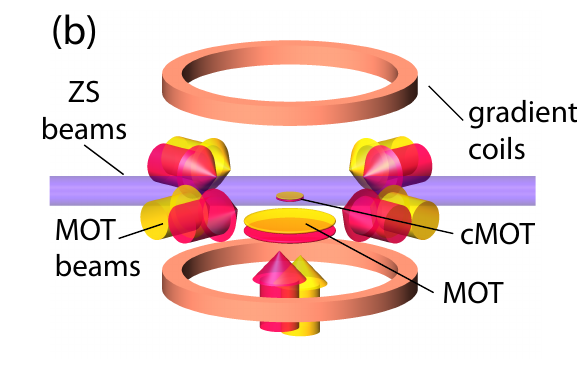}
    }
    \hfill\hfill

    \vspace*{-.0cm}

    \includegraphics{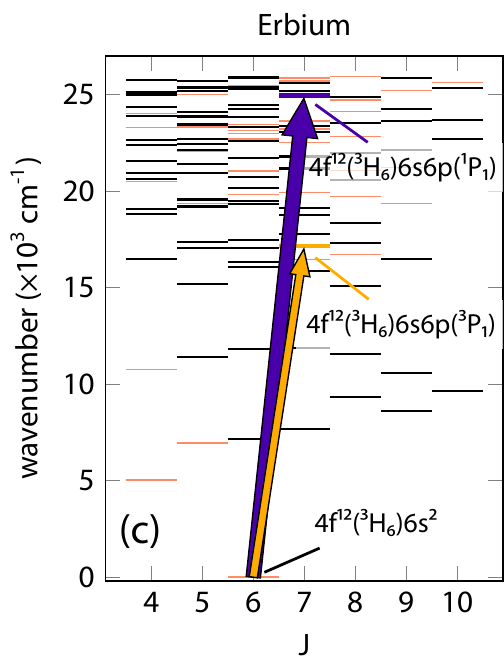}
    \includegraphics{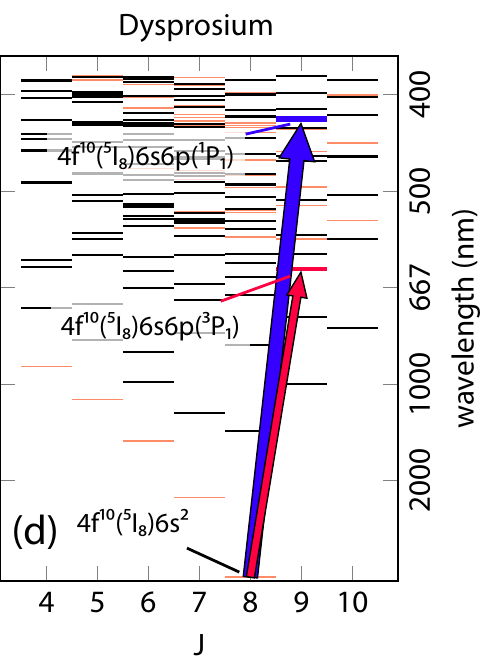}
\hfill
\parbox{7.5cm}{
\vspace*{-6.6cm}

\begin{ruledtabular}
\begin{tabular}{lrr}
\includegraphics[trim = 0 2 0 1,clip]{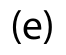}
& Er\hspace*{2em} & Dy\hspace*{2em} \\
\colrule
\begin{tabular}{l}
isotopes\\
abund. (\%)
\end{tabular}& 
\begin{tabular}{rrrr}
166 & 167 & 168 & 170 \\
34 & 23 & 27 & 15
\end{tabular} & 
\begin{tabular}{rrrr}
161 & 162 & 163 & 164 \\
19 & 25 & 25 & 28
\end{tabular} \\[2ex]
$T_\text{melt}$ & \SI{1529}{\celsius} & \SI{1407}{\celsius} \\
$p_\text{vap,\SI{1200}{\celsius}}$ & $\SI{1e-3}{mbar}$ & $\SI{1e-2}{mbar}$ \\
$L,S~(I,F)$ & $5, 1~(\nicefrac{7}{2}, \nicefrac{5..19}{2}$) & $6, 2~(\nicefrac{5}{2}, \nicefrac{11..21}{2}$) \\
$\mu; a_{d} = \smash{\frac{m\mu_0 \mu^2}{4\pi\hbar^2}}$ & $7\,\mu_B; 200\,a_0$ & $10\,\mu_B; 390\,a_0$ \\
$\lambda_\text{broad}$ & \SI{401}{nm}  & \SI{421}{nm} \\
$\Gamma_\text{broad}/2\pi$& \SI{27.5}{MHz} & \SI{32.2}{MHz}\\
$T_\text{D,broad}$ & \SI{659}{\micro\kelvin} & \SI{774}{\micro\kelvin} \\
$\lambda_\text{narrow}$& \SI{583}{nm} &  \SI{626}{nm} \\
$\Gamma_\text{583,626}/2\pi$& \SI{190}{kHz} & \SI{136}{kHz}\\
$T_\text{D,narrow}$ & \SI{4.6}{\micro\kelvin} & \SI{3.2}{\micro\kelvin} \\
$I_\text{sat,583,626}$ & \SI{110}{\micro\watt\per\cm\squared} & \SI{72}{\micro\watt\per\cm\squared}
\end{tabular}
\end{ruledtabular}
}
\caption{Illustration of the vacuum apparatus, including the optical setup for the 5B MOT, and atomic properties of Er and Dy. (a)~The Er-Dy vacuum apparatus, including high-temperature oven, transversal cooling chamber, Zeeman slower (ZS), and MOT chamber. The atomic beam propagates from right to left. The ZS beam is reflected by a metallic mirror in vacuum. (b)~Sketch of the working principle of the open-top MOT. The arrows depict the MOT beams, the blue region indicates the ZS beam. (c\,--\,d)~Energy level diagrams for Er~and Dy~up to \SI{26000}{\per\cm} for different total electronic angular momentum quantum numbers~$J$. States with odd (even) parity are indicated by black (orange) horizontal lines. The arrows show the broad and narrow laser-cooling transitions. (e)~Table listing atomic properties of Er and Dy.}
\label{fig:levelsandsetup}
\end{figure*}

A beneficial factor for combining the multi-valence-electron atoms Er and Dy is their similarity in atomic properties; see table in \myref[ (e)]{fig:levelsandsetup}. They both have several stable isotopes with a high natural abundance~$>14\%$, in total 5 bosonic (\textsuperscript{166}Er, \textsuperscript{168}Er, \textsuperscript{170}Er, \textsuperscript{162}Dy, \textsuperscript{164}Dy) and 3 fermionic (\textsuperscript{167}Er, \textsuperscript{161}Dy, \textsuperscript{163}Dy) isotopes. This isotope variety will allow to prepare ultracold Bose-Bose, Bose-Fermi, and Fermi-Fermi quantum mixtures of Er and Dy. Whereas all bosonic isotopes have zero nuclear spin, the fermionic isotopes possess nuclear spins of $I_\text{Er} = 7/2$ and $I_\text{Dy} = 5/2$, leading to 8 and 6 hyperfine states in the electronic ground state, respectively. Both elements exhibit a rich atomic energy spectrum, arising from their submerged-shell electronic configuration, featuring a [Xe] core, a partially filled inner 4f shell, and a closed outer 6s shell. The electron vacancy in the 4f shell is responsible for the large orbital quantum numbers and the respective high magnetic moments of $7\mu_\text{B}$ and $10\mu_\text{B}$ for Er and Dy.

\Myref[ (c\,--\,d)]{fig:levelsandsetup} shows the electronic levels of Er and Dy for wavenumbers up to \SI{26000}{\per\centi\meter} \cite{NIST_ASD}. While most of the possible transitions are dipole-forbidden, both species offer one particularly broad transition, suitable for laser cooling, to the respective singlet 6s6p state (blue arrows in \myref[ (c\,--d\,)]{fig:levelsandsetup}), which we will label as \emph{broad} in the following. The blue transition light has a wavelength of \SI{401}{nm} (\SI{421}{nm}), and the transition linewidth is $\Gamma_\text{broad}/2\pi= \SI{27.5}{MHz}$ (\SI{32.2}{MHz}) for Er~(Dy)~\cite{Frisch2012, Maier2014}. We use this transition for transversal cooling, Zeeman slowing, and absorption imaging. The laser light, driving the broad transition, is derived from grating-stabilized laser diodes, followed by tapered amplifiers and frequency-doubling cavities. The laser systems emit more than \SI{1}{W} of power each. Both systems are frequency-stabilized using signals from modulation-transfer spectroscopy in hollow-cathode lamps~\cite{Frisch2012}.

Following previous single-species experiments with Yb \cite{Kuwamoto1999}, Er \cite{Frisch2012}, and Dy \cite{Maier2014,Dreon2017}, we produce the MOT using an intercombination line driving the transition from the ground state to the triplet 6s6p state (yellow (red) arrow in \myref[ (c\,--d\,)]{fig:levelsandsetup}) at a wavelength of \SI{583}{nm} (\SI{626}{nm}) in Er (Dy) and a linewidth of $\Gamma_\text{583}/2\pi = $ \SI{190}{kHz} ($\Gamma_\text{626}/2\pi = \SI{136}{kHz}$). The narrow-width character of these transitions leads to conveniently low Doppler temperatures of $T_\text{D,Er} = \SI{4.6}{\micro\kelvin}$ and $T_\text{D,Dy} = \SI{3.2}{\micro\kelvin}$. The laser system for the Er MOT is based on a Raman fiber-amplified diode laser at \SI{1166}{nm} and a single-pass frequency-doubling stage, with an output power above \SI{1.7}{W}. The laser system for the Dy MOT is based on two fiber lasers operating at \SI{1050}{nm} and \SI{1550}{nm}, which are amplified and frequency-converted in a single-pass sum-generation stage, resulting in more than \SI{1.6}{W} of output power. Both MOT laser systems are frequency-stabilized against long-term drifts on a home-built ultra-low expansion cavity via a Pound-Drever-Hall lock \cite{Drever1983} and have linewidths below \SI{100}{kHz}.

The experimental procedure promotes our previously demonstrated single-species MOT approach \cite{Frisch2012} to two-species operation. The very similar strengths and wavelengths of the laser-cooling transitions of Er and Dy, their similar masses and melting points greatly simplify the design of the vacuum apparatus and the experimental procedure for the mixture. \Myref[ (a)]{fig:levelsandsetup} shows the experimental apparatus. Er and Dy atoms are emitted from a single high-temperature oven, consisting of two sections: The first section (effusion cell) typically operates at a temperature of \SI{1100}{\celsius}, the second one (hot lip) operates at \SI{1200}{\celsius}. Three apertures of different diameters, placed inside the oven, geometrically collimate the Er-Dy atomic beam before it enters the transversal-cooling chamber. We operate the transversal cooling resonantly on the broad transitions with total powers of \SI{300}{mW} (\SI{120}{mW}) for Er (Dy) and elliptic waists of approximately $w_\text{horz}=\SI{30}{mm}$ and $w_\text{vert}=\SI{6}{mm}$. We observe that the transversal cooling increases the MOT loading rate by a factor of 10 (6) for Er (Dy). The two-species atomic beam is slowed down to about \SI{5}{m\per s} using a two-species Zeeman slower (ZS) of about \SI{35}{cm} length in spin-flip configuration. The magnetic-field values along the ZS are experimentally optimized for Dy and work equally well for Er. The optimal performance of the ZS has been found for laser powers of \SI{57}{mW} (\SI{121}{mW}) with beam waists of \SI{4}{mm} at a detuning of \SI{-520}{MHz} (\SI{-530}{MHz}) for Er~(Dy).

The slow atoms are then captured into a two-species MOT, operating on the respective intercombination line. Taking advantage of the similar wavelengths, we combine the MOT beams for both species in the same fibers. The MOT light is spectrally broadened utilizing electro-optic modulators (EOM) with resonance frequencies of \SI{139}{kHz} (\SI{102}{kHz}) for Er (Dy), which increases the capture range and thus the number of atoms in the MOT by a factor of 4.5~(2). Optimal loading is reached for peak intensities of the laser beams of $I_{583} = 50I_\text{sat,583}$ and $I_{626} = 160I_\text{sat,626}$, with beam waists of \SI{18}{mm}. A pair of vertical coils creates the magnetic quadrupole field for the MOT of $\nabla B=\SI{4.6}{\gauss\per cm}$. A vertical bias field of $B_0 = \SI{2.9}{\gauss}$ shifts the zero-point of the quadrupole field downwards. Additional coil pairs in the horizontal plane compensate for external magnetic fields.

\looseness=1
We produce and study the Er-Dy MOT using two different beam configurations. In the first one, we use a standard 6B geometry with three pairs of orthogonal retro-reflected beams. For the second configuration, we remove the \mbox{top\,$\rightarrow$\,bottom} MOT beam, resulting in an open-top 5B geometry; see \myref[ (b)]{fig:levelsandsetup}. Although this latter configuration would fail for alkali atoms, we observe a very robust operation for our lanthanoid mixture. As discussed in Refs.\,\cite{Kuwamoto1999,Frisch2012,Dreon2017}, the combined effects of the intercombination line and gravity yield a peculiar semi-shell-shaped MOT with the center lying below the magnetic-field zero of the quadrupole field. As a consequence, the atoms mainly absorb photons from the bottom-top $\sigma^-$-polarized beam, whereas the top-bottom beam plays a minor role. Additionally, the predominant absorption of $\sigma^-$ light leads to a spin-polarization of the sample into the lowest Zeeman sublevel \cite{Frisch2012,Dreon2017}. In our 5B approach this effect is pushed to the limit of removing the sixth beam. We note that early MOT experiments on alkali atoms have explored MOT geometries with five non-orthogonally intersecting beams (e.\,g.~at an angle of 120\textdegree) \cite{Arlt1998,diStefano1999}, constituting a very different scenario.

In the following, we will describe and compare the performances of the 6B- and 5B-MOT configuration. The data we present here refer to the mixture of \textsuperscript{168}Er and \textsuperscript{164}Dy. We are also able to trap and cool all other abundant bosonic isotopes with equally good performance, whereas the \textsuperscript{170}Er MOT has smaller numbers, as expected from its low natural abundance. For future studies of the fermionic isotopes, no changes to the experimental apparatus are necessary. In all mixture combinations, we do not observe any mutual influence of one species on the other. We believe that the narrow-line character of the MOT transitions favors this absence of cross-talk.

\begin{figure}[t]
    \includegraphics{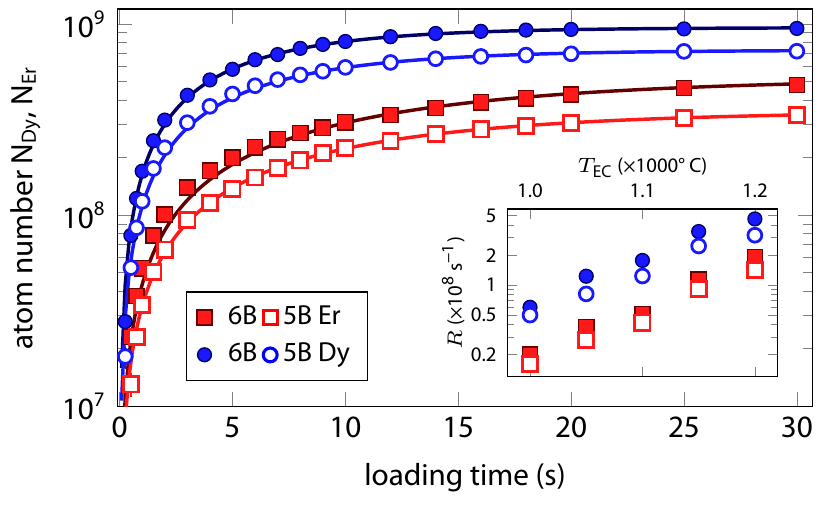}
    \caption{Loading curve of the two-species MOT for 6B and 5B configuration. Full/empty squares (circles) show the data for the 6B/5B Er (Dy) MOT. The corresponding lines are fits to the data, as detailed in the text. The fit parameters are listed in \mytabref{tab:loadingfitvalues}. (inset) Loading rate of the two species in dependence of the effusion cell temperature~$T_\text{EC}$. Note that the hot lip is always kept at $T_\text{HL} = T_\text{EC} + \SI{100}{\celsius}$.}
    \label{fig:loading}
\end{figure}

In a first set of experiments, we study the loading of our two-species MOT in both 6B and 5B configuration. For all atom numbers we report in this letter, we apply a compression phase after the MOT loading and detect the atom numbers using absorption imaging, as described later. \Myref{fig:loading} shows the Er and Dy atom numbers as a function of the MOT loading time. From a fit to the data using a standard loading function $N(t) = N_\text{ss} \cdot \left(1-e^{-\gamma t}\right)$, we extract the loading rate $R$, and decay rate $\gamma$, with the steady-state atom number $N_\text{ss} = R/\gamma$; see \Mytabref{tab:loadingfitvalues}. In both 6B and 5B configuration, we observe a very efficient loading of the two-species MOT. After about \SI{10}{s} of loading, the atom numbers approach their steady state value of some \num{e8} atoms, see \Mytabref{tab:loadingfitvalues}. The difference in the loading curves between Er and Dy is due to their different vapor pressures. We remarkably find that the performance of the 5B MOT is only slightly below the one of the 6B configuration. Moreover, even our 5B double-species MOT shows atom numbers similar or larger than the ones previously reported for single-species Er or Dy MOT experiments \cite{Berglund2007,Frisch2012,Lu2010,Maier2014,Dreon2017}. Judging from our experience with Er \cite{Frisch2012}, we are thus confident that these numbers are sufficient for reaching quantum degeneracy.

The ratio of vapor pressures of Er and Dy is typically larger than 10 at the same temperature \cite{Habermann1964} and strongly temperature-dependent, which would prevent an efficient simultaneous MOT loading. To compensate for this, we selectively heat up the atoms exploiting the two-section design of the oven. We fill the effusion cell with a 33/67\,\%~alloy of Er/Dy  and the hot-lip section with pure Er. We operate the oven with a differential temperature  of \SI{100}{\celsius} between the two sections. In the temperature range from \SIrange{1000}{1200}{\celsius} for the first section, we expect to reduce the vapor-pressure ratio among the two species to about 2.3. We investigate this effect by repeating the loading experiments for different temperatures of the effusion cell, while keeping the hot lip always \SI{100}{\celsius} hotter, see \myref[~(inset)]{fig:loading}. We observe a roughly constant loading ratio between 2.5 to 3.5, which confirms the above expectations and shows that our concept of differential heating works very efficiently.

\begin{table}[h]
\begin{ruledtabular}
\caption{Loading rates $R$, decay rates $\gamma$, and steady-state atom numbers $N_\text{ss}$ as obtained from fits to the data shown in \myref{fig:loading}. Given are the values for both species in 5B and 6B configuration. Also listed are the lifetimes obtained from the data in \myref{fig:lifetime}.
\label{tab:loadingfitvalues}}
\begin{tabular}{lllllll}
& \multicolumn{2}{c}{Er\qquad\qquad} & \multicolumn{2}{c}{Dy\qquad\qquad} \\
  & 5B & 6B & 5B & 6B \\
  \colrule \\[-2ex]
  $R$ ($\times \SI{e8}{\per\second}$)& \num{0.35(1)} & \num{0.45(2)} &
     \num{1.21(4)} & \num{1.79(3)} \\
  $\gamma~(s^{-1})$ & 0.100(2) & 0.086(6) & 0.166(7) & 0.187(5) \\
  $N_\text{ss}~(\times \num{e8})$ & \num{3.5(1)} & \num{4.5(3)} & \num{7.3(4)} & \num{9.6(3)} \\
  lifetime cMOT (ms) & \num{515(65)} & \num{475(50)} & \num{374(44)} & \num{345(23)}\\
\end{tabular}
\end{ruledtabular}
\end{table}

\begin{figure}[ht]
    \includegraphics{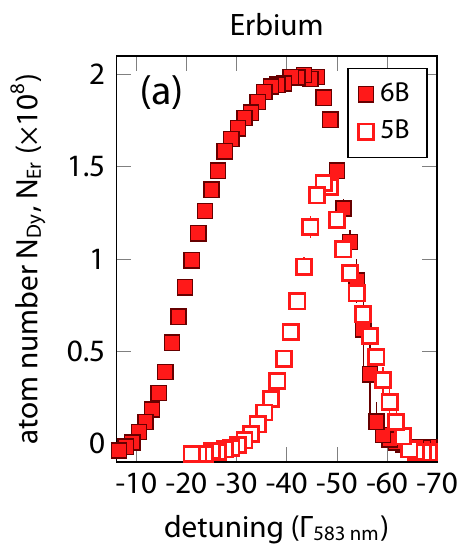}
    \includegraphics{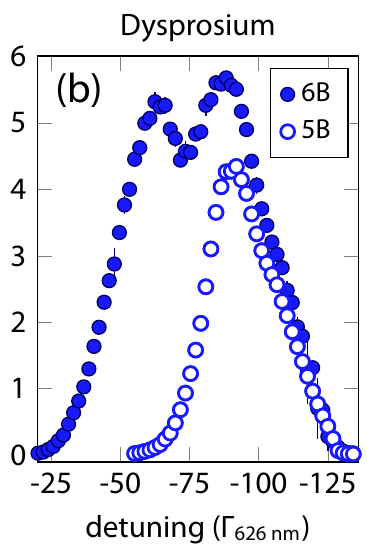}%
    \caption{Dependence of atom numbers in the cMOT on the initial MOT detuning, in units of the respective linewidth of the narrow transition of $\Gamma_{583} = 2\pi\times\SI{190}{kHz}$ ($\Gamma_{626} = 2\pi\times\SI{136}{kHz}$) for Er (Dy). Both species show broad ranges of detuning for efficient MOT loading in 6B (the dip in Dy numbers is explained in the text), while the 5B shows a narrower range. The optimal detunings are nearly equal for 5B and 6B configuration.}
    \label{fig:detuning}
\end{figure}

In a second set of experiments, we systematically study the effect of the MOT-light detuning from the respective resonant atomic frequency on the atom number and compare the results for the 6B and 5B configuration after \SI{5}{s} of loading; see \myref{fig:detuning}. For both species, we see a clear rise of atom numbers with increasing detuning. After reaching a maximum value the numbers undergo a sharp decrease for large detunings. This decrease can be simply explained by the spatial downshift of the MOT position with increasing detuning, eventually causing the atoms leaving the recapture volume. Here, the equal behavior of the 6B and 5B MOT indicates that the top-bottom beam does not play a significant role. At intermediate detunings, however, the two configurations clearly show a different behavior. In particular, the 6B MOT has a much broader range of operation than the 5B configuration. We believe that this difference is due to the fact that the central cloud position approaches the magnetic-field zero point with decreasing detuning. In absence of the top-bottom beam, atoms above the magnetic-field zero do not experience a restoring light force towards the trap center and may escape from the MOT. Contrariwise, in the 6B approach, these atoms are retrapped, resulting in the broader operation range in terms of detuning. \mbox{We attribute} the dip in Dy atom numbers to the spatial overlap of the MOT with the ZS beam in this detuning range, which causes off-resonant scattering and increased losses. For Er, this effect is not observed.

\begin{figure}[ht]
    \includegraphics{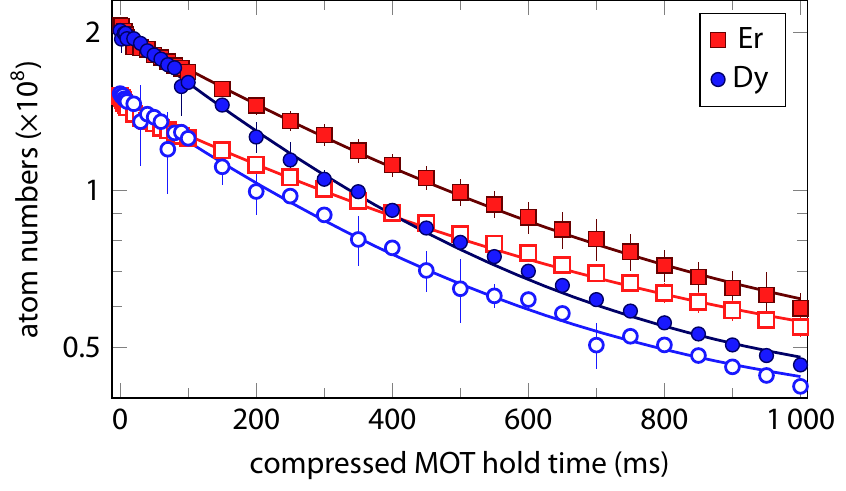}
    \caption{Lifetime of the cMOT for both Er (red) and Dy~(blue) in 5B (open symbols) and 6B (filled symbols) configuration. We adjust the MOT loading time to compare samples with equal atom numbers. Solid lines show exponential fits to the respective data. From the fits we extract lifetimes of 
    \SI{515+-65}{ms}/\SI{475+-50}{ms} (\SI{374+-44}{ms}/\SI{345+-23}{ms}) for Er (Dy) in the 5B/6B cMOT.
    }
    \label{fig:lifetime}
\end{figure}

Finally, we study the lifetime of the compressed MOT (cMOT). The compression phase is essential to efficiently load an optical dipole trap (ODT) in future experiments, as the compression reduces the temperature and increases the density of the mixture. After loading the MOT, we switch off the spectral broadening EOMs, the ZS light, and block the atomic beam with a mechanical shutter. The compression has a duration of \SI{200}{ms}, in which we (i) reduce the detuning of the MOT light to $10\,\Gamma_{583}$ ($18\,\Gamma_{626}$), (ii) decrease the MOT-beam power to $I_{583} = 0.17 I_\text{sat,583}$ and $I_{626} = 0.6 I_\text{sat,626}$, (iii) ramp down the magnetic field gradient to $\nabla B = \SI{4.3}{\gauss\per\centi\meter}$, and (iv) switch off the vertical bias magnetic field. As shown in \myref{fig:lifetime}, we observe that the double-species cMOT has a very long lifetime. The results of fits to the data are listed in \Myref{tab:loadingfitvalues}. Additionally, we extract temperatures of \SI{11(1)}{\micro \kelvin} (\SI{10(1)}{\micro \kelvin}) for Er (Dy) in both the 5B and 6B configuration from time-of-flight measurements. From our experience, we are certain that the observed lifetimes and temperatures are fully sufficient for an efficient loading into an ODT as the next step towards quantum degeneracy.

In summary, we have demonstrated efficient cooling and trapping of an Er and Dy mixture in a two-species MOT operating on narrow-line transitions. In addition, we demonstrate a new beam geometry for our two-species MOT, which consists of only five laser beams in an open-top orthogonal setting. This geometry has the big advantage to completely free the optical access from the top, which greatly simplifies the implementation of a high resolution imaging, as well as optical lattices. The recorded temperatures and atom numbers provide ideal conditions for subsequent evaporative cooling towards quantum degeneracy. Additionally, our setup is prepared for future implementation of an additional science chamber for the preparation and study of Er and Dy Rydberg atoms. The multi-valence-electron nature of these species will open novel excitation and manipulation schemes using, e.\,g.,~4f shell electrons \cite{Hostetter2015,Dunning2016}.

We acknowledge A.~Frisch, T.~Pyragius, C.~Zhang for support in the initial phase of the experiment. We thank the ERBIUM and the Dy-K teams in Innsbruck, the Er team at Harvard, as well as M.~Sohmen and C.~Politi for fruitful discussions. This work is supported by the ERC Consolidator Grant (RARE, no.~681432) and a FET Proactive project (RySQ, no.~640378) of the EU H2020. G.\,D.~acknowledges support by the Austrian Science Fund FWF within the DK-ALM: W1259-N27.


\input{main.bbl}

\end{document}

%% file: main.bbl
%